\documentclass[aps, prl, amsmath,amssymb, reprint, showpacs,superscriptaddress,groupedaddress, nofootinbib ]{revtex4-1}

\usepackage{graphicx}
\usepackage{dcolumn}
\usepackage{bm}
\usepackage{amsmath}
\usepackage{color}

\usepackage{ulem}
\usepackage{dcolumn}  
\usepackage{amsmath}
\usepackage{multirow}
\usepackage{graphicx}   
\usepackage{subfigure}
\usepackage{comment}
\usepackage{color}
\usepackage[colorlinks,bookmarks=false,citecolor=blue,linkcolor=red,urlcolor=blue]{hyperref}
\usepackage{nicefrac}
\usepackage{soul}
\usepackage{pifont}
\usepackage{gensymb}
\usepackage{upgreek}
\usepackage{amsmath}
\usepackage{graphicx}
\usepackage{dcolumn}
\usepackage{bm}

\begin{document}

\title{Indirect band gap semiconductors for thin-film photovoltaics: High-throughput calculation of phonon-assisted absorption}
\author{Jiban Kangsabanik,$^{1*}$ Mark Kamper Svendsen,$^1$ Alireza Taghizadeh,$^1$ Andrea Crovetto,$^2$ Kristian S. Thygesen$^1$ }
\email{jibka@dtu.dk, thygesen@fysik.dtu.dk}	
\affiliation{$^1$CAMD, Computational Atomic-Scale Materials Design, Department of Physics, Technical University of Denmark, 2800 Kgs. Lyngby Denmark.}
\affiliation{$^2$National Centre for Nano Fabrication and Characterization (DTU Nanolab), Technical University of Denmark, 2800 Kgs. Lyngby Denmark.}

\begin{abstract}
Discovery of high-performance materials remains one of the most active areas in photovoltaics (PV) research. Indirect band gap materials form the largest part of the semiconductor chemical space, but predicting their suitability for PV applications from first principles calculations remains challenging. Here we propose a computationally efficient method to account for phonon assisted absorption across the indirect band gap and use it to screen 127 experimentally known binary semiconductors for their potential as thin film PV absorbers. Using screening descriptors for absorption, carrier transport, and nonradiative recombination, we identify 28 potential candidate materials. The list, which contains 20 indirect band gap semiconductors, comprises both well established (3), emerging (16), and previously unexplored (9) absorber materials. Most of the new compounds are anion rich chalcogenides (TiS$_3$, Ga$_2$Te$_5$) and phosphides (PdP$_2$, CdP$_4$, MgP$_4$, BaP$_3$) containing homoelemental bonds, and represent a new frontier in PV materials research. Our work highlights the previously underexplored potential of indirect band gap materials for optoelectronic thin-film technologies.
\end{abstract}

\maketitle


\section*{I. Introduction}

Renewable electricity produced by photovoltaics (PV) is likely to become a cornerstone of a future sustainable and climate-neutral energy system.
Although the current PV technology is mainly based on silicon (Si) wafer solar cells (26.7$\%$ efficiency)\cite{martin2022efftable, Yoshikawa2017}, thin film technologies have distinct advantages such as mechanical flexibility, reduced material usage, potential for low-cost, low-temperature processing, and integration with existing Si technology in tandem solar cells. After decades of research and device optimization various thin film absorber materials, like GaAs (29.1$\%$), CdTe (22.1$\%$), InP (24.2$\%$), CIGS (23.4$\%$), etc.\cite{wanlass2017systems, green2019pushing, nakamura2019cd, martin2022efftable} have reached efficiencies comparable to the Si-based technology. Most recently, hybrid lead-halide perovskite (MAPbI$_3$) has emerged as an additional PV candidate material with a record efficiency of 25.8$\%$~\cite{Min2021}. 

Although promising, the thin film PV materials still suffer from major drawbacks associated with low material abundance (In, Ga), toxicity (Cd, As, Pb), and long-term device stability under ambient conditions (hybrid perovskites), which hinder their large-scale deployment. As such, the discovery of new (thin-film) PV materials remains a relevant and timely challenge. Within the PV literature, the most widely used selection-metric for absorber materials is the electronic band gap. Indeed, Shockley and Queisser demonstrated that thermodynamic detailed balance sets an upper limit to the power conversion efficiency (PCE) of a semiconductor given solely by its band gap.\cite{shockley1961detailed} More realistic PV descriptors have been developed by invoking further physical properties like the optical absorption coefficient, the carrier mobility, and internal quantum efficiency, along with device related aspects such as film thickness, light-trapping schemes, etc.\cite{tiedje1984limiting, yu2012identification, blank2017selection, kim2020upper}  

Recently, the substantial increase in computer power along with improvements in theoretical methods have ushered in an era where it is possible to screen for new materials faster, cheaper, and more rationally than the traditional approach based on experimental trial and error. In particular, high-throughput density functional theory (DFT) calculations have been employed to identify suitable PV candidate materials.\cite{zhao2017design, kuhar2018high, choudhary2019accelerated, li2019high, luo2021high, dahliah2021high, gan2022robust} Indirect band gap semiconductors, which occupy a major portion of the semiconductor space, have mostly been ignored in these studies. This is because indirect gap materials will absorb photons less efficiently around the band gap, which is likely to lead to low PCE in thin film devices. On the other hand, it has been argued that a small indirect gap might be beneficial for PV applications due to reduced radiative recombination.\cite{hutter2017direct, kirchartz2017decreasing} This situation clearly calls for a clarification. 

The major bottleneck in evaluating the PV performance of an indirect band gap semiconductor from first-principles, is modeling indirect absorption.
This is a higher order process involving one or several phonons (or defects or free carriers) in addition to the photon. Such calculations are very demanding. As a consequence, ab-initio calculations of momentum-indirect absorption has so far been limited to a few materials\cite{noffsinger2012phonon, kang2018first, monserrat2018phonon, menendez2018phonon, bravic2019finite, ha2020boron, ge2020direct, jian2021strained, willis2022prediction}. 

In this work, we propose an approximate method to evaluate phonon-assisted absorption in indirect band gap semiconductors, which only involves the $\Gamma$-point optical phonons and therefore is feasible for high-throughput studies. After establishing the accuracy of the approach for a few prototypical materials, we use it to evaluate the key PV performance parameters for 127 stable, binary semiconductors of which 97 have indirect band gaps. In addition to already known materials, we identify a number of new indirect band gap materials that show promising potential for thin film PV. All the materials along with their calculated properties are available in an open online database. To the best of our knowledge, this is the first high-throughput study evaluating indirect gap materials for PV, which includes calculation of phonon-assisted absorption.

\section*{II. Method}
\begin{figure*}[t!]
	\centering
	\includegraphics[scale=0.59]{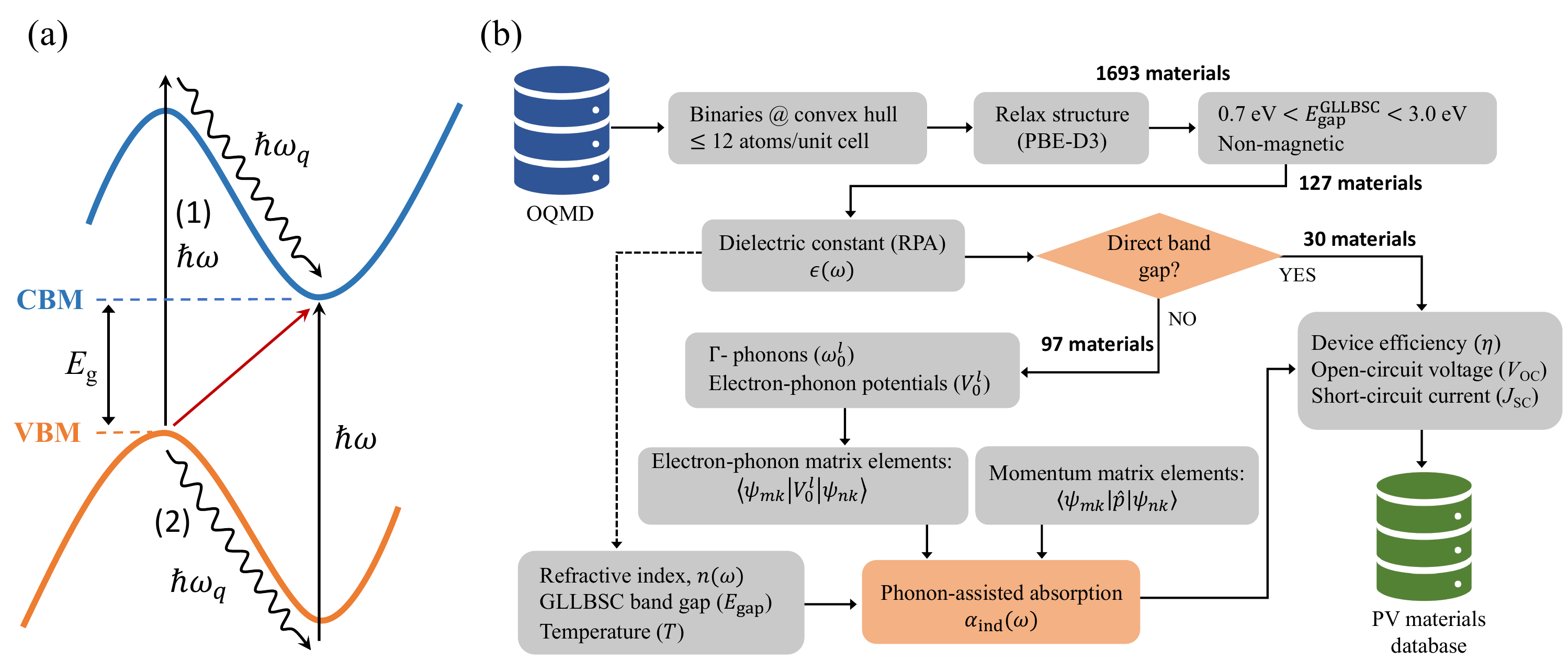}
	\caption{(a) Schematic diagram of phonon-assisted optical absorption. (b) Computational workflow used to evaluate the PV performance parameters for direct and indirect band gap semiconductors.}
	\label{Fig1}
\end{figure*}

Phonon assisted absorption is a second order process, which involves a photon-phonon pair satisfying (crystal) momentum conservation (see Fig.\ref{Fig1}(a)). Because of the involved sum over intermediate states and the need to calculate all the phonons and electron-phonon couplings, its calculation from first-principles is very demanding. In this work we use a simplified approach that involves only the optical $\Gamma$-point phonons. Within second order perturbation theory, the phonon-assisted absorption is given by,

\begin{widetext}
\begin{equation}
    \alpha(\omega)=\frac{4\pi e^2}{m^2\epsilon_0  c}\frac{1}{n(\omega)\omega }\frac{1}{N_k^2} \sum_{\boldsymbol{k}_{1}\boldsymbol{k}_{2},c,v,l} |t^{l,\pm}_{c\boldsymbol{k}_{2}\leftarrow v\boldsymbol{k}_{1}}|^{2}  \delta(E_{c\boldsymbol{k}_{2}}-E_{v\boldsymbol{k}_{1}}-\hbar\omega \pm  \hbar\omega^l_{\boldsymbol{q}}) 
    \label{eqn1}
\end{equation}
\end{widetext}
where,
\begin{widetext}
\begin{equation}
	t^{l,\pm}_{c\boldsymbol{k}_{2}\leftarrow v\boldsymbol{k}_{1}}= \sqrt{n^l_{\boldsymbol{q}}}\Bigg[ \sum_{\alpha}\frac{\langle\psi_{c\boldsymbol{k}_{2}}|V^l(\boldsymbol{q})|\psi_{\alpha\boldsymbol{k}_{1}}\rangle \langle\psi_{\alpha\boldsymbol{k}_{1}}|\hat{\boldsymbol{e}}.\boldsymbol{p}| \psi_{v\boldsymbol{k}_{1}}\rangle}{E_{v\boldsymbol{k}_{1}}-E_{\alpha\boldsymbol{k}_{1}}+\hbar\omega}	+\sum_{\beta}\frac{\langle\psi_{c\boldsymbol{k}_{2}}|\hat{\boldsymbol{e}}.\boldsymbol{p}| \psi_{\beta\boldsymbol{k}_{2}}\rangle \langle\psi_{\beta\boldsymbol{k}_{2}}|V^l(\boldsymbol{q})|\psi_{v\boldsymbol{k}_{1}}\rangle}{E_{v\boldsymbol{k}_{1}}-E_{\beta\boldsymbol{k}_{2}}\pm\hbar\omega^l_{\boldsymbol{q}}}\Bigg]
	\label{eqn2}
\end{equation}
\end{widetext}
Here $\hbar \omega$ is the energy of the incident photon, $l$ denotes the phonon mode and $\mathbf q=\mathbf k_2-\mathbf k_1$ its momentum, and $n(\omega)$ is the refractive index of the semiconductor. The two sums in the square bracket of Eq. (\ref{eqn2}) correspond to the two transition pathways denoted '1' and '2' in Fig.\ref{Fig1}(a). The matrix elements appearing in Eq. (\ref{eqn2}) contain the electron-phonon potential, $V^l(\mathbf q)$, and coupling to a photon field of polarisation $\hat{\mathbf{e}}$, respectively. The phonon occupation factors, $n^l_\mathbf{q}$, are given by the Bose-Einstein distribution. 
\begin{figure*}[t!]
	\centering
	\includegraphics[scale=0.8]{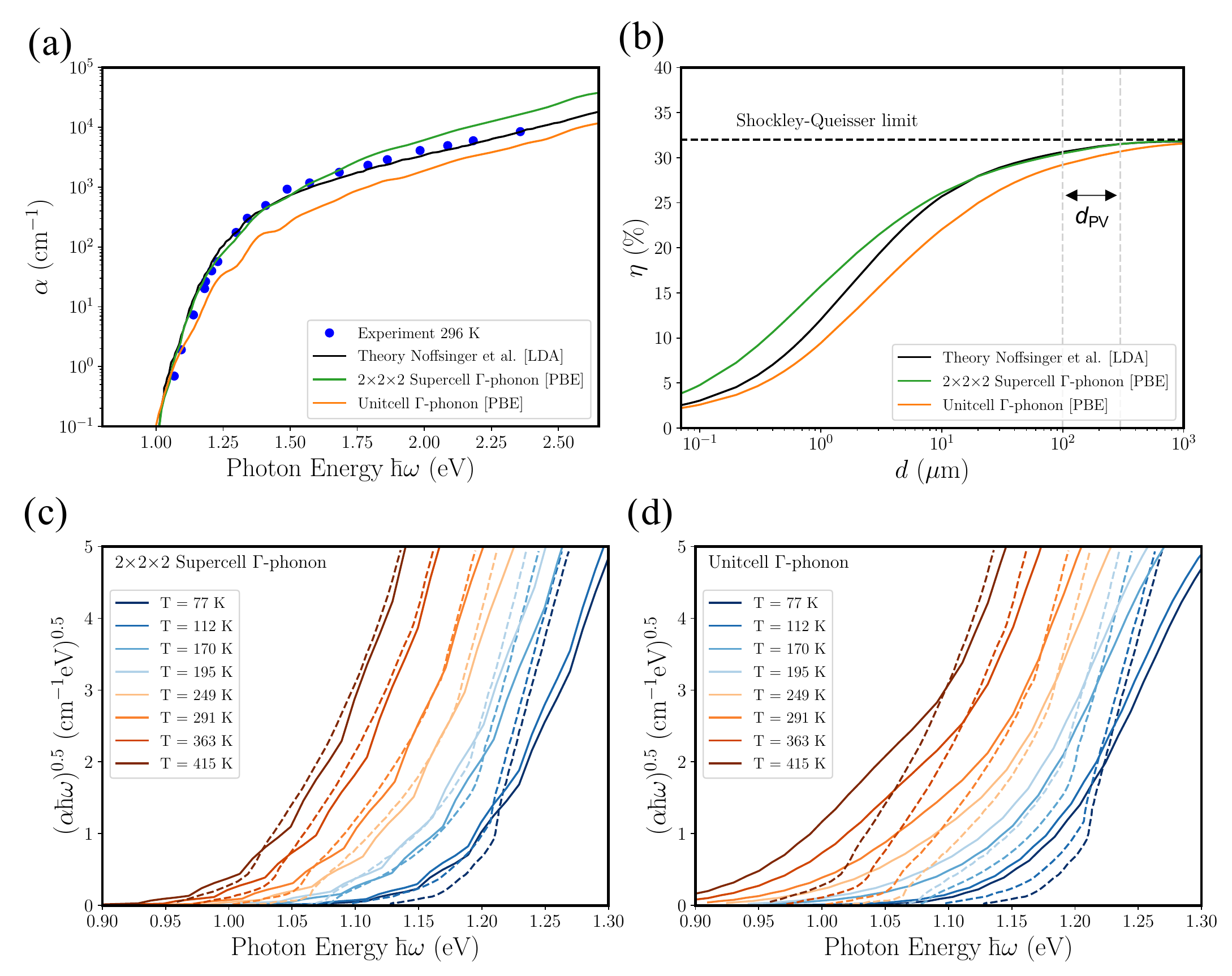}
	\caption{(a) Comparison of phonon assisted absorption in Si, calculated using the approximation Eqs.(\ref{eqn1}-\ref{eqn2}) (green, orange), previous calculations including all phonons (black), and experimental data at T = 296 K (blue). The corresponding Shockley-Queisser device efficiencies as a function of film thickness are shown in (b). Here '$d_{\mathrm{PV}}$' is the usual thickness of the Si active layer in PV cells. (c-d) Comparison of calculated phonon-assisted absorption onsets (solid lines) with experimental data (dashed lines) at different temperatures. In (c) the $\Gamma$-phonons of a $2\times 2 \times 2$ supercell are used in Eqs.(\ref{eqn1}-\ref{eqn2}), while in (d) the $\Gamma$-phonons of the primitive Si unit cell are included. The calculated spectra have been  shifted horizontally to match the experimental absorption onset.}
	\label{Fig2}
\end{figure*}

In our approximation, we assume that scattering on a $(l,\mathbf q)$-phonon can be replaced by scattering on a $(l,\mathbf 0)$-phonon (acoustic phonons are disregarded). Specifically, we keep all transitions in Eq. (\ref{eqn2}), but make the replacements 
\begin{eqnarray}\label{eq:approx1}
\omega_{\mathbf q}^l &\to& \omega_{\mathbf 0}^l\\ \label{eq:approx2}
\langle \psi_{c\mathbf k_1}|V^l(\mathbf k_2-\mathbf k_1)|\psi_{v\mathbf k_1}\rangle  &\to& \langle u_{c\mathbf k_1}|V^l(\mathbf 0)|u_{v\mathbf k_1}\rangle, 
\end{eqnarray}
where $u$ is the periodic part of the Bloch function. Our approximation avoids the computation of the full $\mathbf q$-dependent phonons and electron-phonon couplings and instead requires these quantities only at the $\Gamma$-point. The efficiency gain is largest for materials with small unit cells and it becomes exact in the limit of a large unit cell. In particular, due to band folding, the exact result for any crystal is obtained by using a supercell containing a sufficiently large number of copies of the primitive cell (although this strategy is inefficient as momentum selection rules are not exploited). Below we illustrate the performance of our approach for the case of bulk silicon, and refer to the supplementary information (SI) for further tests (BP, MoS$_2$, MoSe$_2$, etc.), all showing encouraging results. Details of the DFT calculations can be found in section \ref{methodology}.
 
 Fig. \ref{Fig2}(a) shows the phonon-assisted absorption in silicon obtained within our approximate scheme applied to the Si primitive unit cell (orange - 3 optical phonons included) and a supercell containing $2\times 2 \times 2$ repetitions of the primitive cell (green - 45 optical phonons). Our results are compared to experimental data ($T= 296$K) and a previous ab-initio result including all phonons. The theoretical spectra have been shifted horisontally to match the experimental absorption onset. We have checked that the use of the LDA and PBE xc-functionals lead to insignificant changes in the spectrum. While our $2\times 2\times 2$ supercell $\Gamma$-phonon approach shows excellent agreement with both sets of reference data, the primitive unit cell $\Gamma$-phonon approach slightly underestimates the absorption coefficient.
 Next, we add the first order absorption coefficient, $\alpha_{\text{dir}}$, (calculated in the random phase approximation (RPA)) to the indirect absorption coefficients and 
 evaluate the radiative device PCE, $\upeta$, as a function of thickness of the Si absorber, see Fig. \ref{Fig2}(b). Overall, the different descriptions of the indirect absorption leads to similar PCEs with the most approximate $\Gamma$-phonon description underestimating the PCE by maximum $5\%$ relative to experiments. The last two panels of Fig. \ref{Fig2} show the temperature dependence of the indirect absorption as calculated with the $\Gamma$-phonons of the $2\times 2 \times 2$ supercell (c) and primitive cell (d), respectively. The supercell calculations are in excellent agreement with the experimental data (dashed lines), except for small deviations at temperatures below 100K. The primitive cell calculations show larger deviations -- also at higher $T$ close to the absorption onset.


The workflow used to calculate the PV performance parameters
is illustrated in Fig. \ref{Fig1}(b). For 1693 thermodynamically stable elemental and binary compounds from the Open Quantum Materials
Database (OQMD) with up to 12 atoms/unit cell, we perform structural relaxations using the PBE\cite{perdew1996generalized} xc-functional with D3 corrections\cite{grimme2010consistent} to account for van der Waals interactions (more details can be found in computational details section). The data set includes 338 semiconductors\cite{svendsen2022computational} for which we evaluate the band gaps using
both the PBE and the GLLBSC\cite{kuisma2010kohn} xc-functional. Keeping only nonmagnetic compounds with GLLBSC band gap between 0.7 eV and 3.0 eV, reduces the data set further down to 127 compounds. For these materials, the complex dielectric function is calculated in the RPA (this is a first order response function including only direct momentum-conserving transitions with local field effects accounted for). For the 97 semiconductors with indirect band gap, the phonon assisted absorption coefficient is calculated using the $\Gamma$-phonon approximation in the primitive unit cell, c.f. Eqs. (\ref{eqn1}-\ref{eqn2}). Finally, the PV parameters are evaluated from the total absorption coefficient, including first and second order processes, as described in SI in section S1.

\section*{III. Results and discussion}

\begin{figure*}[t!]
	\centering
	\includegraphics[scale=0.59]{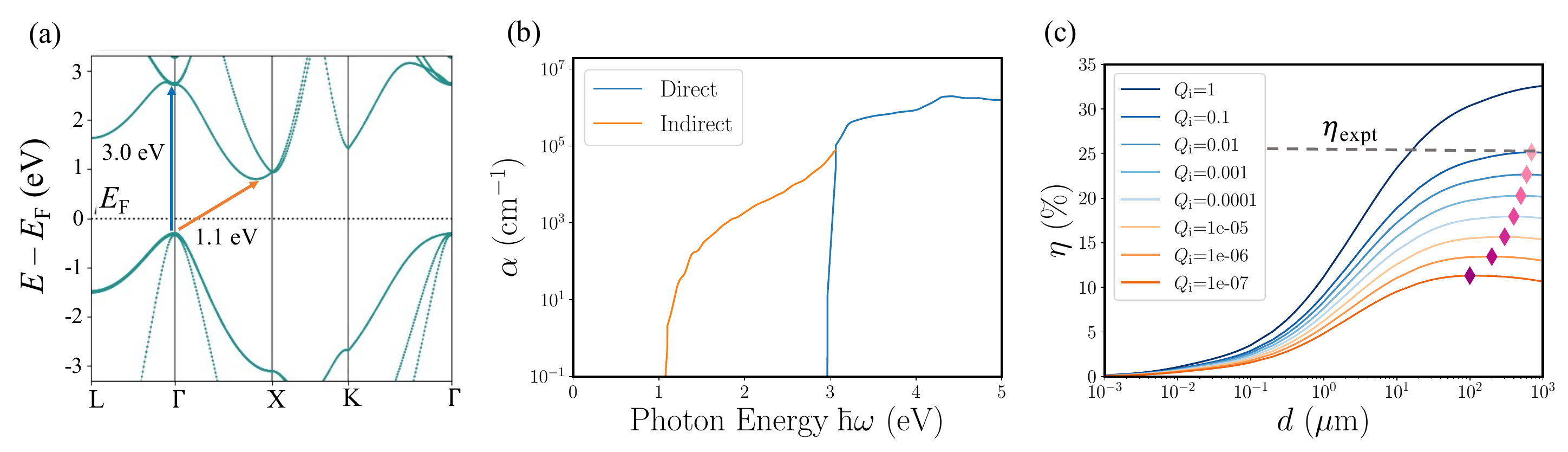}
	\caption{(a) Electronic band structure of Silicon (Si), (b) Total optical absorption coefficient for Si (including direct and phonon assisted (indirect) absorption). (c) Efficiency ($\upeta$) vs. thickness ($d$) plot for various internal (luminescence) quantum efficiencies ($Q_\text{i}$). Maximum $\upeta$ at each $Q_\text{i}$ is denoted using $\diamond$ symbol. Here experimental efficiency $\upeta_\text{expt}$ for a typical Si-solar cell is denoted by black dashed line.}
	\label{Fig3}
\end{figure*}

\subsection*{A. Descriptors}
The main descriptor used to quantify the potential of a PV absorber material, is its power conversion efficiency (PCE), $\upeta$, which can be expressed as, 
\begin{equation}
	\upeta =\frac{\max(V[J_\text{sc}-J_{r}^\text{tot}(e^{\frac{eV}{k_\text{B}T}}-1)])}{\int_{0}^{\infty}E I_\text{sun}(E) dE}
	\label{eqPVmain}
	\end{equation}

In this expression $V$ is the voltage, $J_\text{sc}$ is the short-circuit current density, $J_{r}^{\text{tot}}$ is total recombination current density, and $T$ is the temperature. In the denominator, we have the input power density, where $I_\text{sun}(E)$ represents the photon flux from the incident solar spectrum (we have used the standard AM1.5G solar spectrum). More details can be found in Methodology section in SI. Here we stress that the only input required to obtain $\upeta$ from Eq. (\ref{eqPVmain}) in the radiative limit (i.e. when non-radiative recombination processes are neglected and thus internal quantum efficiency $Q_i=1$) is the material's absorption spectrum. 

In the simplest theory, the absorbance is assumed to be zero below the band gap and one above. In the radiative limit this yields the well known Shockley-Queisser (SQ) limit.\cite{shockley1961detailed} Clearly, this description fails to differentiate between different materials with the same band gap. While this is a reasonable approximation for bulk cells, a realistic description of thin film devices must account for the finite thickness and frequency dependent absorption.\cite{tiedje1984limiting, yu2012identification, blank2017selection}

A popular PV performance descriptor that invokes the absorption coefficient is the Spectroscopic Limited Maximum Efficiency (SLME).\cite{yu2012identification} The SLME sets $J_r^{\mathrm{tot}}=J_{\mathrm{rad}}\exp(\Delta E_\text{{g}}^\text{dir}/k_\text{B}T)$ in Eq. (\ref{eqPVmain}) to account for non-radiative recombination. In other words, the effect of an indirect band gap is taken into account via an increased non-radiative loss rate. In fact, this can lead to erroneous conclusions for indirect band gap materials. For example, the calculated SLME for Si at 100 $\upmu$m thickness is only 1$\%$,  while the experimental PCE is around 25\%. We argue that explicitly accounting for indirect absorption processes is essential for a proper description of PV performance parameters across all band gap types. In particular, given the exact absorption coefficient, Eq. (\ref{eqPVmain}) would yield the exact PCE in the radiative limit. More detailed discussion is provided in the SI section S3.
To account for non-radiative recombination we use the formalism proposed by Blank \emph{et al.}\cite{blank2017selection} as discussed later.

\subsection*{B. Si as a case study}
Before discussing our high-throughput results, we illustrate the PV analysis in detail for the well known case of Si. The silicon crystal has an indirect band gap of 1.1 eV, whereas the direct band gap at the $\Gamma$-point is 3.0 eV (see Fig. \ref{Fig3}(a)). Since the first-order absorption coefficient (e.g. evaluated using the independent particle approximation, the RPA, or the Bethe-Salpeter Equation), only includes vertical transitions, it will fail to account for photon absorption below the direct gap. The effect of including phonon-assisted transitions in the absorption coefficient can be seen in Fig. \ref{Fig3}(b) (the orange curve is the same as shown in Fig. 2(a)). While the second-order processes are weak, they are completely essential for a correct description of the opto-electronic properties. In the radiative limit we obtain a PCE of 31.3\% (with phonon-assisted processes) and 4.9\% (only first-order processes) for a 100 $\upmu$m thick Si film.

To move beyond the radiative approximation one must account for non-radiative recombination such as Shockley-Read-Hall recombination\cite{shockley1952statistics, hall1952electron} and Auger processes\cite{huldt1971band}. These are challenging to evaluate from first principles and to some extent dependent on the sample quality via the concentration and types of defects in the material. In addition, there are additional loss channels that depend on the device architecture, such as surface and interface recombination. Here we will neglect the latter types, and only consider the non-radiative processes taking place within the absorbing layer. To that end, we follow Blank \emph{et al.}\cite{blank2017selection} who proposed a detailed balance compatible approach where non-radiative processes are included via the internal luminescence quantum efficiency, $Q_\text{i}$ (the ratio of the radiative to the total recombination rates inside the material under equilibrium conditions), which becomes a free parameter, see Section S1 in SI for more details. 

\begin{figure*}[t!]
	\centering
	\includegraphics[scale=0.8]{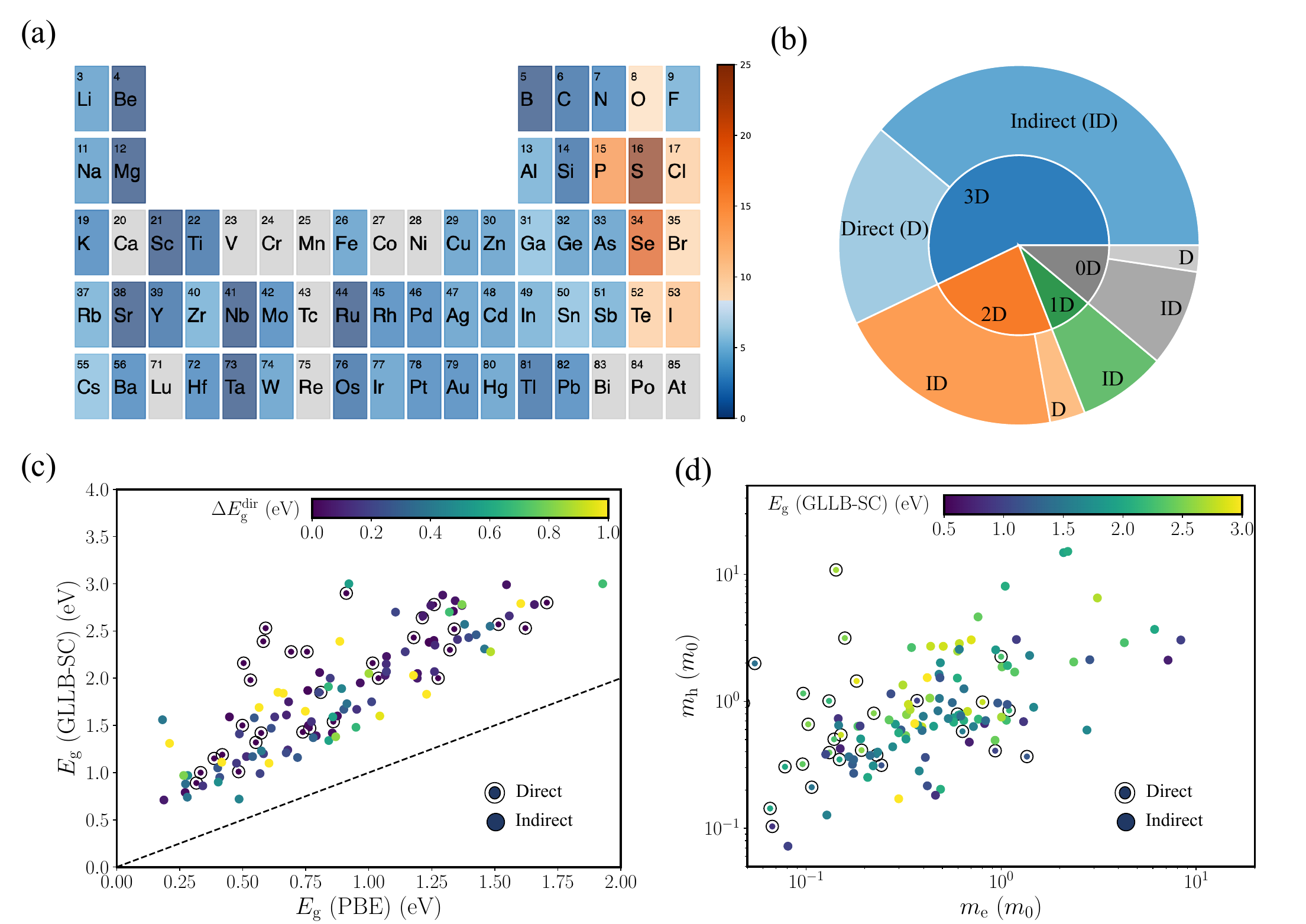}
	\caption{(a) Elemental distribution of the dataset in the periodic table. Color scale shows number of occurrence of an element in the final dataset. Grey colored boxes represents element which are absent in the dataset. (b) Dimensionality distribution of the dataset as 3D, 2D, 1D, and 0D bulk materials. Associated indirect (ID) and direct (D) gap material distributions are shown in the outer circle. (c) GLLB-SC vs PBE band gap distribution for the selected materials. Color scale represents difference between direct and indirect band gap ($\Delta E_\text{{g}}^\text{{dir}}$). (d) Effective mass (in units of electron rest mass m$_0$) distribution of the compounds. Here the colors scale shows corresponding GLLB-SC bandgaps.  }
	\label{Fig4}
\end{figure*}

Fig. \ref{Fig3}(c) shows the calculated $\upeta$ vs film thickness for different values of $Q_\text{i}$. As expected, with increasing non-radiative recombination the maximum achievable efficiency decreases. Perhaps less obvious, also the optimal thickness (thickness at which $\upeta$ is maximum) decreases when the non-raditaive recombination increases. This effect occurs because the optimal thickness occurs exactly when the increase in absorption due to a further increase in thickness is balanced by the increase in the non-radiative recombination. The existence of an optimal thickness then follows by noting that the former decays exponentially through the material while the latter is constant. We obtain an efficiency close to the experimental value ($\upeta_\text{expt}$) of 25$\%$ at optimal film thickness for $Q_\text{i}$ around $10^{-1}$, see Fig. \ref{Fig3}(c).
The experimental value of $Q_\text{i}$ in the highest-efficiency Si solar cell~\cite{Yoshikawa2017} is $1.6 \times 10^{-2}$~\cite{green2019pushing}, which implies a maximum efficiency of about 23\%. We ascribe this small underestimation of the PCE to the $\Gamma$-point phonon approximation and to the fact that the front surface of a real Si solar cell is textured to promote light scattering and increase the average optical path of photons in Si. Both approximations lead to a slight underestimation of the fraction of incident light absorbed in a film of given thickness, see Fig. 2(b). 

\subsection*{C. High-throughput screening}
Moving on to the main purpose of our work, namely the identification of high-performance indirect band gap PV absorbers, Fig. \ref{Fig4}(a) shows the distribution of chemical elements in the 127 semiconductors that were subject to the workflow in Fig. 1(b). There is a relatively homogeneous distribution of cations, with the exception of certain 3d transition metals, which are absent because we have excluded magnetic materials. The anionic elements are dominated by the chalcogens (in particular sulphur) followed by the halogens. We note that we did not filter out materials containing rare or toxic elements at this step. In Fig. \ref{Fig4}(b) we show how the 127 compounds are distributed over the dimensionality of their covalently bonded atomic clusters. The dimensionality analysis follows the method of Larsen \emph{et al.}\cite{larsen2019definition}. It can be seen that the materials are almost evenly divided between '3D' and 'non-3D' materials. About 50\% of the latter class are van der Waals (vdW) layered 2D materials while the remainder are vdW-bonded 1D or 0D clusters. For each dimensionality class we show the fraction of direct (D) and indirect (ID) band gaps.

\begin{figure*}[t!]
	\centering
	\includegraphics[scale=0.7]{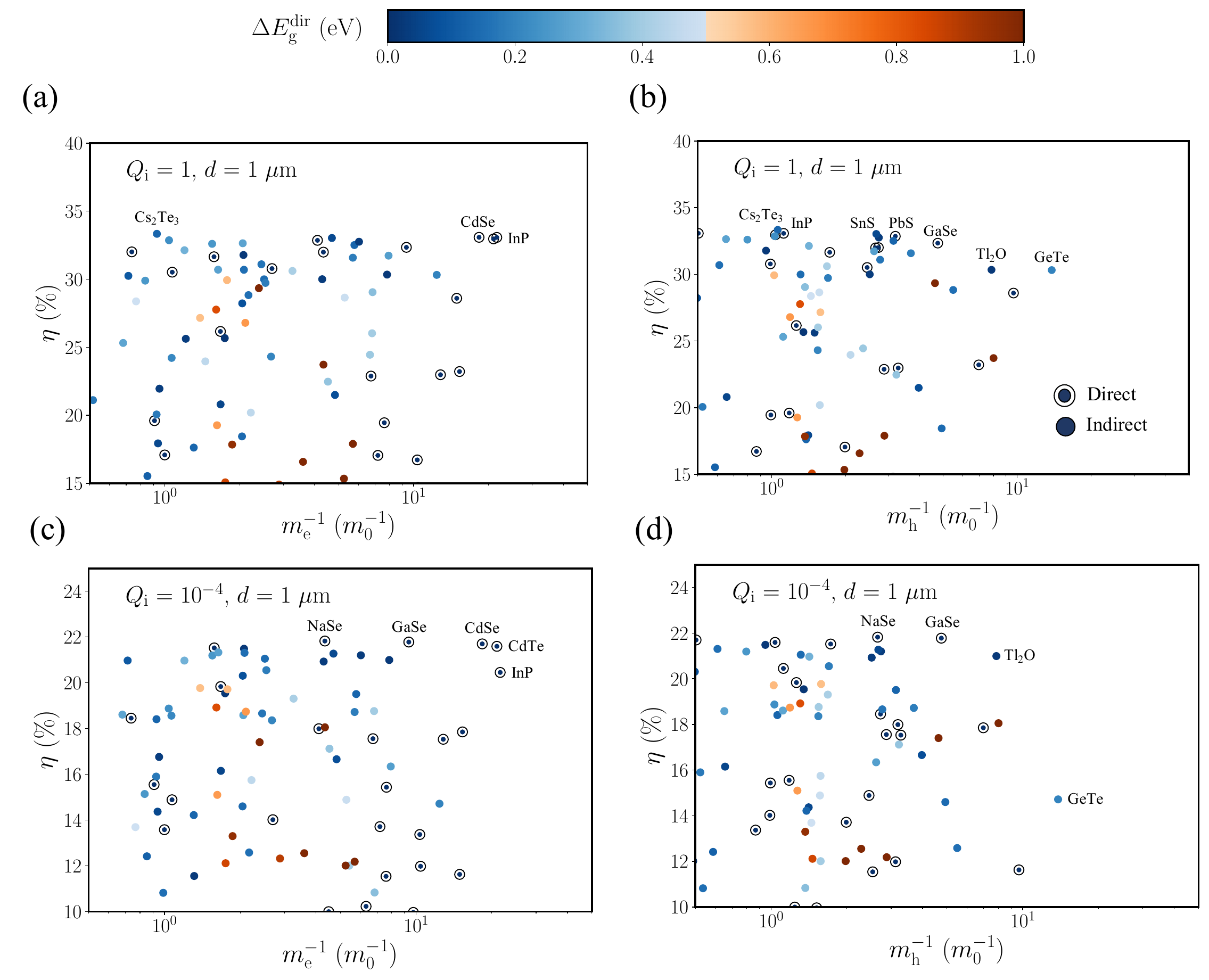}
	\caption{Pareto distribution for the selected photovoltaic (PV) materials (device efficiency ($\upeta$) vs inverse effective masses ($m_\text{e}^{-1}$, $m_\text{h}^{-1}$)). (a) Radiative (Q$_\text{i}$=1) $\upeta$ vs. $m_\text{e}^{-1}$, (b) Radiative (Q$_\text{i}$=1) $\upeta$ vs. $m_\text{h}^{-1}$. (c) $\upeta$ (at $Q_\text{i}$=10$^{-4}$) vs. $m_\text{e}^{-1}$, (d) $\upeta$ (at $Q_\text{i}$=10$^{-4}$) vs. $m_\text{h}^{-1}$. Color scale represents difference between direct gap and fundamental electronic band gap ($\Delta E_\text{{g}}^\text{{dir}}$). $\upeta$ is calculated at $T$= 298 K, and film thickness $d$= 1 $\upmu$m.}
	\label{Fig5}
\end{figure*}

In total, more than 75$\%$ of the 127 semiconductors possess an indirect band gap. Fig. \ref{Fig4}(c) shows the band gap distribution for our dataset. It is well known that the PBE xc-functional systematically underestimates the band gap. The GLLB-SC functional provides more accurate band gaps at little extra computational cost\cite{kuisma2010kohn,rasmussen2015computational}. An important reason behind the good performance of the GLLB-SC is includes a correction of the Kohn-Sham band gap to account for the so-called derivative discontinuity. The GLLB-SC band gaps range from 0.7 eV to 3.0 eV, whereas with PBE band gap lies between 0.2 eV and 2 eV. From the color scale it is clear that most of the indirect gap materials have $\Delta E_\text{{g}}^\text{{dir}}$ $<$ 0.5 eV while a only few compounds have $\Delta E_\text{{g}}^\text{{dir}}$ $>$ 1.0 eV (all materials with $\Delta E_\text{{g}}^\text{{dir}}$ above 1 eV have yellow color code) . 

High absorption and low non-radiative recombination rates do not guarantee good PV performance. As an additional criterion, the carrier mobility, $\upmu=\frac{e\tau}{m^{*}}$, should be high enough to ensure efficient electron-hole separation and extraction of charges at the contacts. The relaxation time, $\tau$, accounts for scattering on other charge carriers, phonons, and crystal imperfections. While it can be accurately estimated from first principles\cite{kaasbjerg2012phonon,kaasbjerg2020atomistic}, its calculation remains challenging in high-throughput studies. As a consequence we focus on the carrier effective mass 
\begin{equation}
    m^*=3[\frac{1}{m_{xx}^*}+ \frac{1}{m_{yy}^*}+ \frac{1}{m_{zz}^*}]^{-1},
\end{equation} 
where $m_{ii}^*$ is the effective mass along direction $i$, which we calculate by fitting a parabola to the band extrema. Fig. \ref{Fig4}(d) gives an overview of the hole and electron effective masses in the 127 semiconductors considered. The band gap is indicated by the color code.

\begin{figure*}[t!]
	\centering
	\includegraphics[scale=0.64]{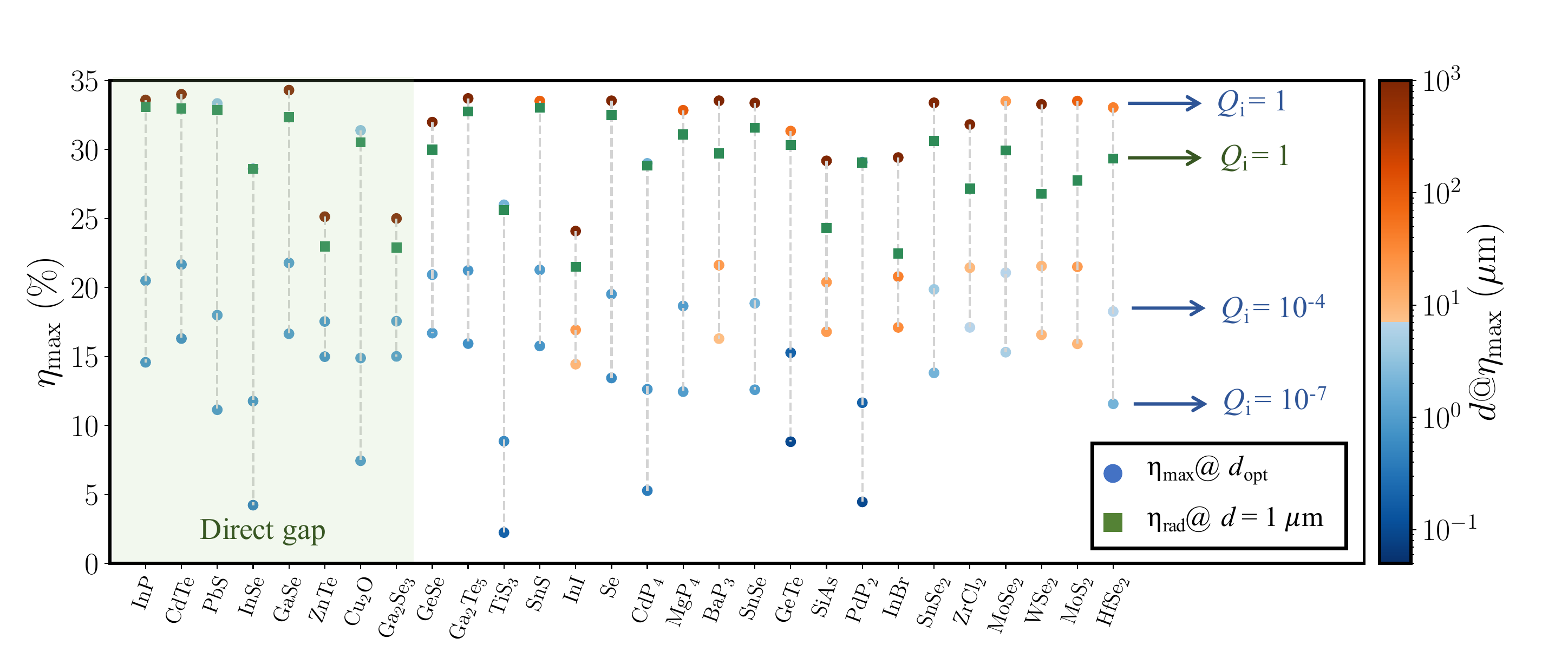}
	\caption{Maximum radiative photovoltaic device efficiency ($\upeta_{\text{max}}$) as a function of different internal quantum efficiencies (log$_{10}$Q$_{\text{i}}$ = 0, -4 and -7) for the selected compounds ($\upeta_{\text{max}}$ $>$ 20$\%$ at $Q_{\text{i}}$ = 1, $d$= 1 $\upmu m$, $T$ = 298 K,   $m_{\text{e}}$, $m_{\text{h}}$ $<$ $m_0$). Color scale represents corresponding film thickness (d) at $\upeta_{\text{max}}$. Additionally, $\upeta_{\text{rad}}$ [$\upeta$ at $Q_{\text{i}}$=1] at $d$= 1 $\upmu$m is shown using $\square$ symbol. Light green shaded region represents the direct band gap materials.}
	\label{Fig6}
\end{figure*}

To evaluate the potential of a given PV material taking both optical absorption and carrier transport into account, we plot in Fig. \ref{Fig5} the PCE (for film thickness $d$= 1 $\upmu m$) against the inverse carrier effective mass of electrons (left) and holes (right). We label all materials that are "Pareto optimal"\cite{khoroshiltseva2016pareto}, i.e. for which no other material has higher values for both $\upeta$ and $1/m^*$. In the upper panels (a-b), $\upeta$ is evaluated in the radiative limit, $Q_i=1$, while we use a more realistic value of $Q_i=10^{-4}$ in the lower panels. The color scale  represents the difference in direct and indirect band gap, $\Delta E_\text{{g}}^\text{{dir}}$.

Some well known thin film PV materials, like CdTe, PbS and InP, as well as emerging materials, like GaSe, SnS are found on the Pareto fronts in Fig. \ref{Fig5}. Not unexpectedly, most of the promising PV materials have $\Delta E_\text{{g}}^\text{{dir}}$ $<$ 0.5 eV (blueish rather than reddish color). In Table I, we list all materials satisfying the criteria: $\upeta$$_{\text{max}}$ $>$ 20\% at $Q_\text{i}$ = 1, $d$= 1 $\upmu$m, $T$ = 298 K, and $m_\text{e}$, $m_\text{h}$ $<$ $m_\text{0}$.    The materials are ordered according to the size of $\Delta E_\text{{g}}^\text{{dir}}$. While there is a tendency for smaller $\Delta E_\text{{g}}^\text{{dir}}$ to correlate with higher $\upeta$, there are also clear deviations from this trend. For example, SnSe$_2$ with $\Delta E_\text{{g}}^\text{{dir}}=0.41$ eV has $\upeta=30.61\%$, which is not far from the highest value $\upeta=33.07\%$ obtained for the direct band gap material InP.

\begin{table*}[t]
\caption{\label{tab:table_xfinal}
Various relevant parameters (Compound, space group, Dimensionality (dim), Band gap ($E_\text{g}$) (calculated using GLLB-SC functional), difference between direct and indirect gap ($\Delta E_\text{{g}}^\text{{dir}}$), effective masses ($m_\text{e}$ and $m_\text{h}$), device efficiencies ($\upeta_1$ [$d$=1 $\upmu$m, $Q_\text{i}$=1] and $\upeta_2$ [$d$=1 $\upmu$m, $Q_\text{i}$=10$^{-4}$]), ICSD-id, etc. ) for the selected compounds for thin film PV technology.}
\begin{ruledtabular}
\textbf{Established}\\[2pt]
\begin{tabular}{cccccccccc}
Formula$^\text{[Ref]}$ & space group & dim & $E_\text{g}$ & $\Delta E_\text{{g}}^\text{{dir}}$(eV) & $m_\text{e}$(m$_0$) & $m_\text{h}$(m$_0$) & $\upeta_1$(\%) & $\upeta_2$(\%) & ICSD-id\\[2pt]
\hline

InP$^\text{{\cite{wanlass2017systems}}}$ & F-43m & 3D & 1.32 & 0.00 &0.05 & 0.89 & 33.07 & 20.45 & 41443 \\[2pt]
CdTe$^\text{\cite{martin2022efftable}}$ & F-43m & 3D & 1.43 & 0.00 &0.05 & 0.97 & 32.96 & 21.59 & 31844 \\[2pt]
PbS$^\text{\cite{choi2020cascade}}$ & Fm-3m & 0D & 1.15 & 0.00 &0.24 & 0.31 & 32.86 & 17.99 & 62192 \\[2pt]

\end{tabular}
\\[2pt]
\textbf{Emerging, PCE  $>$ 5\%}\\[2pt]
\begin{tabular}{cccccccccc}
Formula$^\text{[Ref]}$ & space group & dim & $E_\text{g}$ & $\Delta E_\text{{g}}^\text{{dir}}$(eV) & $m_\text{e}$(m$_0$) & $m_\text{h}$(m$_0$) & $\upeta_1$(\%) & $\upeta_2$(\%) & ICSD-id\\[2pt]
\hline
InSe$^\text{\cite{MartinezPastor1987}}$ & P6$_3$/mmc & 2D & 0.89 & 0.00 &0.07 & 0.10 & 28.60 & 11.63 & 30377 \\[2pt]
Cu$_2$O$^\text{\cite{Minami2016}}$ & Pn-3m & 3D & 1.01 & 0.00 &0.93 & 0.41 & 30.52 & 14.89 & 261853 \\[2pt]
Se$^\text{\cite{todorov2017ultrathin}}$ & P3$_1$21 & 1D & 1.24 & 0.12 & 0.17 & 0.32 & 32.51 & 19.51 & 164271 \\[2pt]
WSe$_2$ $^\text{\cite{Kim2022a}}$ & P6$_3$/mmc & 2D & 1.48 & 0.66 & 0.47 & 0.84 & 26.80 & 18.74 & 40752 \\[2pt]
\end{tabular}
\\[2pt]
\textbf{Emerging, PCE  $<$ 5\%}\\[2pt]
\begin{tabular}{cccccccccc}
Formula$^\text{[Ref]}$ & space group & dim & $E_\text{g}$ & $\Delta E_\text{{g}}^\text{{dir}}$(eV) & $m_\text{e}$(m$_0$) & $m_\text{h}$(m$_0$) & $\upeta_1$(\%) & $\upeta_2$(\%) & ICSD-id\\[2pt]
\hline
GaSe$^\text{\cite{ishikawa2021photovoltaic}}$ & P6$_3$/mmc & 2D & 1.47 & 0.00 &0.11 & 0.21 & 32.34 & 21.78 & 41978 \\[2pt]
ZnTe$^\text{\cite{Wang2011c}}$ & P3$_1$ & 3D & 2.00 & 0.00 &0.08 & 0.31 & 22.98 & 17.53 & 80076 \\[2pt]
Ga$_2$Se$_3$ $^\text{\cite{ho2020ga2se3}}$ & Cc & 3D & 2.00 & 0.00 &0.15 & 0.35 & 22.88 & 17.56 & 35028 \\[2pt]
GeSe$^\text{\cite{liu2021boosting}}$ & Pnma & 2D & 1.60 & 0.03 & 0.23 & 0.40 & 30.00 & 20.93 & 17006 \\[2pt]
SnS$^\text{\cite{sinsermsuksakul2014overcoming}}$ & Pnma & 2D & 1.39 & 0.06 & 0.21 & 0.37 & 33.03 & 21.27 & 24376 \\[2pt]
InI$^\text{\cite{Dunlap-Shohl2018}}$ & Cmcm & 0D & 2.00 & 0.07 & 0.21 & 0.25 & 21.49 & 16.66 & 38129 \\[2pt]
SnSe$^\text{\cite{nandi2022vapor}}$ & Pnma & 2D & 1.20 & 0.18 & 0.18 & 0.27 & 31.58 & 18.72 & 16997 \\[2pt]
GeTe$^\text{\cite{jones2016design}}$ & R3m & 2D & 0.99 & 0.20 & 0.08 & 0.07 & 30.32 & 14.72 & 655497 \\[2pt]
SnSe$_2$ $^\text{\cite{yu2012snse}}$ & P-3m1 & 2D & 1.23 & 0.41 & 0.31 & 0.60 & 30.61 & 19.31 & 43594 \\[2pt]
MoSe$_2$ $^\text{\cite{memaran2015pronounced}}$ & P6$_3$/mmc & 2D & 1.34 & 0.58 & 0.56 & 0.98 & 29.93 & 19.72 & 49800 \\[2pt]
MoS$_2$ $^\text{\cite{wi2014enhancement}}$ & P6$_3$/mmc & 2D & 1.38 & 0.83 & 0.62 & 0.77 & 27.77 & 18.92 & 24000 \\[2pt]
HfSe$_2$ $^\text{\cite{afzal2021fast}}$ & P-3m1 & 2D & 1.11 & 1.02 & 0.42 & 0.22 & 29.34 & 17.40 & 638902 \\[2pt]

\end{tabular}

\textbf{Unexplored}\\[2pt]
\begin{tabular}{cccccccccc}
Formula & space group & dim & $E_\text{g}$ & $\Delta E_\text{{g}}^\text{{dir}}$(eV) & $m_\text{e}$(m$_0$) & $m_\text{h}$(m$_0$) & $\upeta_1$(\%) & $\upeta_2$(\%) & ICSD-id\\[2pt]
\hline
Ga$_2$Te$_5$ & I4/m & 3D & 1.42 & 0.03 & 0.17 & 0.37 & 32.75 & 21.20 & 1085 \\[2pt]
TiS$_3$ & P2$_1$/m & 2D & 0.79 & 0.03 & 0.82 & 0.67 & 25.62 & 8.72 & 42072 \\[2pt]
CdP$_4$ & P2$_1$/c & 3D & 0.86 & 0.15 & 0.46 & 0.18 & 28.83 & 12.58 & 25605 \\[2pt]
MgP$_4$ & P2$_1$/c & 3D & 1.10 & 0.17 & 0.41 & 0.36 & 31.10 & 18.66 & 113 \\[2pt]
BaP$_3$ & C2/m & 3D & 1.41 & 0.18 & 0.39 & 0.59 & 29.73 & 20.55 & 23618 \\[2pt]
SiAs & C2/m & 2D & 1.75 & 0.20 & 0.37 & 0.65 & 24.31 & 18.36 & 43227 \\[2pt]
PdP$_2$ & C2/c & 3D & 0.88 & 0.36 & 0.15 & 0.73 & 29.05 & 10.83 & 166275 \\[2pt]
InBr & Cmcm & 0D & 1.73 & 0.37 & 0.22 & 0.31 & 22.47 & 17.12 & 55188 \\[2pt]
ZrCl$_2$ & R3m & 2D & 1.59 & 0.57 & 0.72 & 0.63 & 27.16 & 19.77 & 20144 \\[2pt]

\end{tabular}

\end{ruledtabular}
\end{table*}

In Fig. \ref{Fig6}, we show how $\upeta$ varies with the amount of non-radiative recombination for some of the most promising materials. Here we vary $Q_\text{i}$ from 1 (radiative limit) to 10$^{-7}$ (dominant non-radiative recombination) and plot the corresponding maximum efficiency, $\upeta_{\text{max}}$, and optimal film thickness, $d@\upeta_{\text{max}}$, indicated by the color code. We also show $\upeta_{\text{rad}}$ at 1 $\mu$m film thickness. In the SI we provide more detailed plots of $\upeta$ versus film thickness for various values for 
$Q_\text{i}$ (as in Fig. \ref{Fig3}(c)) for all the materials in Fig. \ref{Fig6}. The order of the materials in Fig. \ref{Fig6} is such that $\Delta E_\text{{g}}^\text{{dir}}$ increases from left to right, i.e. direct band gap materials, indicated by green shaded region, are found to the left. Considering only $\upeta_\text{rad}$ when comparing a direct and indirect gap compound is not sufficient, as $Q_\text{i}$ can be different for different materials. For example, CdTe has $\upeta_\text{SQ}$ $\approx 33\%$, whereas the known highest device efficiency is 22.1$\%$.\cite{martin2022efftable} This efficiency is slightly higher than $\upeta_{\text{max}}$ at $Q_\text{i}$=10$^{-4}$. The measured $Q_\text{i}$ on the highest-efficiency CdTe solar cell is $8 \times 10^{-3}$~\cite{green2019pushing}, indicating that some additional loss mechanisms are present besides non-radiative recombination. At the same time, we can see a number of indirect gap materials which show $\upeta_{\text{max}} > 20\%$, for practical $Q_\text{i}$ (10$^{-2}$-10$^{-4}$) ranges. The optimal device thickness in case of indirect materials are higher mainly due to low absorption at the indirect gap. We can see a trend that with increasing $\Delta E_\text{{g}}^\text{{dir}}$ the optimal device thickness (at different $Q_\text{i}$ values) increases, which means with larger $\Delta E_\text{{g}}^\text{{dir}}$ the suitability for thin film PV technology might reduce. Still, there are a number of indirect gap compounds showing excellent $\upeta_{\text{max}}$ at thin film thickness range. These materials are discussed in the next section.

There has been a lot of discussion whether a small indirect gap can be beneficial for PV application.\cite{hutter2017direct, kirchartz2017decreasing} With this study, we clarify how an indirect and direct gap material compare in the radiative limit for small film thicknesses (see SI section S7 for a detailed discussion). We find that at smaller film thicknesses ($d=1\upmu$m), in the radiative limit (high mobility) adding a small indirect gap to a direct gap material will never show higher $\upeta_\text{rad}$, as the gain in absorption (higher $J_\text{sc}$) will be counteracted by higher radiative loss (lower $V_\text{oc}$). However, an indirect gap material with
small $\Delta E_\text{{g}}^\text{{dir}}$ can exhibit higher $\upeta_\text{rad}$ compared to a direct gap material with same fundamental gap, as long as the gap is below the SQ maximum (1.1 eV).  For example, an indirect gap material with $E_\text{{g}}$ = 0.95 eV and $\Delta E_\text{{g}}^\text{{dir}}$=0.1 eV can in principle show higher $\upeta_\text{rad}$ compared to a material with direct $E_\text{{g}}$ of 0.95 eV.  This is mainly due to the opposite effect (higher gain in $V_\text{oc}$ compared to loss in $J_\text{sc}$) as discussed for the previous case. With this study, using our formalism we can provide real material examples here. For example, InSe is a direct gap material with $E_\text{{g}}$=0.89 eV showing $\upeta_\text{rad}$ of 28.60\%. In comparison, CdP$_4$, which has an indirect gap of 0.86 eV and direct gap of 1.01 eV, shows higher $\upeta_\text{rad}$ of 28.83\% (see Table I). More examples can be found in SI section S7 and related tables.

\subsection*{D. Discussion of final list of materials}

Among the materials that pass all screening criteria, we discard the ones that have been shown to be unstable in air (elemental I, Cs$_2$P$_3$, Rb$_2$P$_3$, RbSb$_2$, KSb$_2$, Na$_3$Sb, NaSe, YN). We also leave out the Tl-containing materials (Tl$_2$O, Tl$_2$Te$_3$) due to serious toxicity concerns. Finally, we discard the Os- and Ru-containing materials (OsAs$_2$, OsP$_2$, RuP$_2$) due to their very low earth abundance.

The final list of screened materials is presented in Table 1, grouped by maturity level. We note that all of the compounds have an ICSD-id, which means that they have been previously synthesized. CdTe, InP, and PbS are established materials which have been particularly successful in single-junction, multijunction, and quantum dot solar cells, respectively. 

Among the emerging materials identified by our work, there are a few with record PCE above 5\%. They are InSe~\cite{MartinezPastor1987}, Cu$_2$O~\cite{Minami2016}, elemental Se\cite{todorov2017ultrathin}, and WSe$_2$~\cite{Kim2022a}. The other emerging materials have been investigated for PV applications, but their record PCE is below 5\% or they have not yet been incorporated in a working solar cell. Except for InI~\cite{Dunlap-Shohl2018}, all materials in this group are chalcogenides, and the large majority of them have 2D (layered) structures. Group IV, Ga-based, and transition-metal chalcogenides are particularly well represented. In many cases, the development of these layered compounds for PV applications has focused on extremely thin films, to take advantage of the enhanced light absorption that many of these compounds experience in atomically-thin form. ZnTe is a well known hole contact layer in CdTe PV technology, but it has also been tested as a wide-gap photoabsorber for tandem cell applications, although efficiencies are still very low~\cite{Wang2011c}.

The "rediscovery" of many already known PV materials lends credibility to our approach. We note that a few binary compounds with notable PV efficiencies, like Sb$_2$S$_3$ and Zn$_3$P$_2$, are missed by our approach because they have more than 12 atoms in their unit cell. We miss GaAs due to underestimation band gap using GLLB-SC functional (0.68 eV). We also miss CdSe because the calculated hole effective mass is higher than $m_0$. 

We find nine compounds that have not been previously studied as PV materials. We label them "unexplored" in Table 1. To the best of our knowledge, none of these compounds have been synthesized as thin films, except for InBr~\cite{Brothers1990}. Intriguingly, none of these compounds belongs to traditional semiconductor families. What most of these semiconductors have in common is the existence of homoelement bonds, i.e., bonds between the same elements. This feature is absent in the currently known compound semiconductors for PV (III-V, II-VI, perovskites, and derivatives thereof). Due to this qualitative difference, the properties of these semiconductors with hybrid (homo- and heteroelement) bonding cannot be readily extrapolated from the properties of well-known semiconductors with only heteroelement bonds. Thus, these materials could be a new frontier of PV materials discovery.

Four compounds (CdP$_4$, MgP$_4$, BaP$_3$, and PdP$_2$) are phosphorus-rich phosphides, containing metal-P as well as P-P bonds. Due to these additional covalent bonds, phosphorus is in a higher oxidation state (ranging from -1 to -0.5) than the classic -3 state found in optoelectronic phosphides, e.g., InP, Zn$_3$P$_2$, and the II-IV-P$_2$ ternaries. In general, phosphorus-rich phosphides have not been studied as potential PV materials before. They had hardly been grown in thin-film form until a very recent report of single-phase CuP$_2$ films in the P2$_1$/c structure by reactive sputtering and soft annealing in an inert atmosphere~\cite{doi:10.1021/jacs.2c04868}.

In a similar spirit, TiS$_3$ can be classified as a sulfur-rich sulfide due to the presence of both metal-S and S-S bonds. TiS$_3$ has only been minimally investigated in photoelectrochemical cells and not in thin-film form~\cite{Gorochov1983}. In SiAs, all Si atoms are bonded to 3 As atoms and 1 Si atom, so Si can be considered as in the +3 oxidation state, thus explaining the 1:1 stoichiometry. There is one report of SiAs in a photoelectrochemical cell, but the sample was not in thin-film form~\cite{lewerenz1983photoeffect}. In Ga$_2$Te$_5$, 20\% of the Te atoms are only bonded to other 4 Te atoms (Te-Te bonds), and the remaining 80\% is bonded to one Te atom and 2 Ga atoms.

On the other hand, the two halide compounds InBr and ZrCl$_2$ do not have homoelement bonds and are quite different from each other. InBr should be qualitatively similar to the already known InI~\cite{Dunlap-Shohl2018}, but with a larger $\Delta E_\text{{g}}^\text{{dir}}$ and a band gap closer to the SQ optimum. ZrCl$_2$ is one of the first transition metal halides to be identified for potential PV applications. It has the same structure as 3R-MoS$_2$ with Zr in the unusual +2 oxidation state. Unlike most of the known PV semiconductors, ZrCl$_2$ has a lone valence band. It also has a large $\Delta E_\text{{g}}^\text{{dir}}$ and reasonably low effective masses along its 2D layers.


Considering a combination of crustal abundance and elasticity of production~\cite{Vesborg2012}, we estimate that only five compounds from the "unexplored" family could potentially be scaled to terawatt levels. These are TiS$_3$, MgP$_4$, BaP$_3$, SiAs, and ZrCl$_2$. These compounds deserve further investigation.

\section*{IV. Conclusion}

In summary, we propose an efficient methodology to calculate phonon assisted absorption, which enables us to accurately evaluate optical absorption for indirect band gap materials. Using this method, along with various relevant photovoltaic (PV) descriptors (such as device efficiency, effective masses, internal luminescence quantum efficiency, etc.) we evaluate the potential of 127 experimentally known binary semiconductors (of which 97 are have indirect band gap) for thin film PV. After screening the materials based on the PV descriptors and considering toxicity, material abundance and air stability, we identify 28 promising materials of which 20 have an indirect gap. This list includes well established PV materials like InP, CdTe, PbS, as well as emerging ones such as, elemental Se, SnS, SnSe, SnSe$_2$, lending credibility to our approach. Additionally, we discover 9 previously unexplored compounds with potential for thin film PV. Most of these compounds show homoelemental bonds and the constituent elements show deviation from their well known oxidation states. These materials (and their respective stoichiometric classes) have scarcely, if ever, been explored for photovoltaics and could become a new frontier of next generation PV materials research. Further theoretical work on these materials should address their defect tolerance, carrier mobilities and lifetimes. Such studies can benefit from our open database containing the calculated atomic and electronic structures as well as basic PV descriptors for the 127 materials investigated in this work.

\section*{V. Methodology}\label{methodology}
\subsection*{A. Computational Details}
First principles calculations are performed using Density Functional Theory (DFT)\cite{kohn1965self} within the Projector-Augmented Wave (PAW) formalism\cite{blochl1994projector} as implemented in GPAW code\cite{mortensen2005real, enkovaara2010electronic}, in combination with Atomic Simulation Environment (ASE)\cite{larsen2017atomic}.  A plane wave cutoff energy of 800 eV and k-mesh density of 8 \r{A}$^{-1}$ are employed for geometry optimization. Electronic structure calculations are performed using Perdew-Burke-Ernzerhof (PBE)\cite{perdew1996generalized} exchange correlation functional with double zeta polarized (dzp) basis set\cite{larsen2009localized}. The accuracy of dzp basis set was tested in a previous study and reported to produce phonon modes with fair accuracy \cite{taghizadeh2020library}.  The phonon modes are evaluated via calculating the dynamical matrices within the harmonic approximation. Force constant matrices are calculated using small displacement approach\cite{alfe2009phon} as implemented in ASE. Electron-phonon matrix elements are evaluated based on calculation of gradients of effective potential with finite difference technique \cite{kaasbjerg2012phonon} as implemented in GPAW. As our approximation involves only zone-center ($\Gamma$ point) phonons, we use primitive unit cells for respective calculations. Next, the momentum matrix elements are obtained using the finite difference technique as implemented in GPAW. A real-space grid spacing of 0.2 Å is chosen for these calculations. We have used 16$\times$16$\times$16 (8 \r{A}$^{-1}$) k-mesh (density) for systems with less (more) than 4 atoms in the unit cell, when evaluating the electron-phonon coupling matrix elements. For phonon calculations,  8 \r{A}$^{-1}$ (6 \r{A}$^{-1}$) k-mesh density is used for unit cells with less (more) than 4 atoms and the forces are converged until $10^{-7}$ eV/\r{A}. The width of the Fermi-Dirac smearing is kept at 50 meV. While calculating the probability of transition, the Dirac delta function is replaced with a Gaussian function with a smearing width of 40 meV,  which accounts for inhomogeneous spectral broadening. The choice of PBE exchange-correlation functional which is known to slightly overestimate the lattice parameters and significantly underestimate the band gap values, is expected to have some effect on the calculated absorption coefficient and efficiency. In order to account for the band gap underestimation, we have calculated the band gaps with GLLB-SC functional\cite{kuisma2010kohn}, which is known to produce accurate band gaps\cite{castelli2015new} and is highly affordable computationally. During the calculation of probability of transition, the conduction bands are scissor shifted to match the GLLB-SC gap. The effect of temperature is incorporated only into the phonon occupation numbers. We have considered $T$= 298 K for all the calculations. Additional finite temperature effects (e.g. lattice expansion and electron-phonon renormalization, etc.) may exhibit effects on the absorption coefficient and as such device efficiency slightly.  Optical absorption coefficient beyond the direct gap is obtained via calculating the frequency dependent dielectric function within the Random Phase Approximation (RPA) as implemented in GPAW. The computational workflow was constructed using the Atomic Simulation
Recipes (ASR)\cite{gjerding2021atomic} and executed using the MyQueue\cite{mortensen2020myqueue} task scheduler frontend.

\section{Acknowledgements}
We acknowledge funding from the European Research Council (ERC) under the European Union’s Horizon 2020 research and innovation program Grant No. 773122 (LIMA) and Grant agreement No. 951786 (NOMAD CoE). K. S. T. is a Villum Investigator supported by VILLUM FONDEN (grant no. 37789).


\providecommand{\noopsort}[1]{}\providecommand{\singleletter}[1]{#1}%

\end{document}